\documentstyle[epsfig,12pt]{article}
\def \bea{\begin{eqnarray}}
\def \beq{\begin{equation}}

\def \bra#1{\langle #1 |}

\def \eea{\end{eqnarray}}
\def \eeq{\end{equation}}

\def \ket#1{| #1 \rangle}

\def \ob{\overline{B}^0}

\def \pr{\parallel}
\def \ras{\rho_{A_1}^2}
\def \rfas{\rho_{F_A}^2}
\def \rfvs{\rho_{F_V}^2}
\def \rs{\rho^2}

\begin{document}
\Large
\centerline {\bf Factorization in Color-Favored}
\centerline{\bf $B$-Meson Decays to Charm
\footnote{Enrico Fermi Institute preprint EFI 01-02, hep-ph/0101089.
Submitted to Physical Review D.}}
\normalsize
 
\vskip 2.0cm
\centerline {Zumin Luo~\footnote{zuminluo@midway.uchicago.edu} and
Jonathan L. Rosner~\footnote{rosner@hep.uchicago.edu}}
\centerline {\it Enrico Fermi Institute and Department of Physics}
\centerline{\it University of Chicago, 5640 S. Ellis Avenue, Chicago, IL 60637}
\vskip 4.0cm
 
\begin{quote}

Improved $B$ meson decay data have permitted more incisive tests of
factorization predictions.  A concurrent benefit is the ability to constrain
the Cabibbo-Kobayashi-Maskawa matrix element $|V_{cb}|$.  Using a simultaneous
fit to differential distributions $d \Gamma(\ob \to D^{(*)+} l^-
\bar \nu_l)/ dq^2$ and the rates for the color-favored decays $\ob \to D^{(*)+}
(\pi^-,\rho^-,a_1^-)$, we find $|V_{cb}| = 0.0415 \pm 0.0022$.
The slope of the universal Isgur-Wise form factor is described by a
parameter found to be $\rs = 1.52 \pm 0.11$.  Taking the $D_s$ meson
decay constant from the world average of direct
measurements, we predict satisfactorily the branching ratios for $\ob \to
D^{(*)+} D_s^{(*)-}$.  Ratios of helicity amplitudes for color-favored
processes are also found to be in accord with predictions.

\end{quote}
\bigskip

\noindent
PACS Categories:  13.25.Hw, 14.40.Nd, 14.65.Fy, 12.39.Hg

\vfill
\newpage

\section{Introduction}

Semileptonic weak hadron decays provide useful information on form
factors of the weak current.  The lepton pair can then be replaced with a
hadron, permitting the calculation of nonleptonic decays.  Although this hadron
can re-interact with the rest of the system, the effects of this re-interaction
sometimes can be neglected or evaluated.  In such cases one is employing the
{\it factorization hypothesis}.  An early version of this hypothesis \cite{BJ}
was recently justified for certain decays of hadrons containing heavy quarks
\cite{BBNS}.

In the present paper we update and test some factorization predictions first
made a number of years ago \cite{JRFM}.  We compare values of the
Cabibbo-Kobayashi-Maskawa matrix element $|V_{cb}|$ and form factor shapes
obtained from (1) the differential distribution $d \Gamma(\ob \to
D^{(*)+} l^- \bar \nu_l)/dq^2$ \cite{CLEOVcb,ALEPH,DELPHI,OPAL,LEP,CLEOBD}
and (2) the color-favored two-body nonleptonic decays $\ob \to D^{(*)+}
(\pi^-,\rho^-,a_1^-)$.  We find that consistency between nonleptonic and
semileptonic determinations is at least as good as that among the semileptonic
determinations themselves.

Using a combined fit to semileptonic and nonleptonic decays and
the measured value of the $D_s$ meson decay constant $f_{D_s}$, we then
predict the rates for $\ob \to D^{(*)+} D_s^{(*)-}$ and find that they
are in accord with experiment.  We thus find that factorization holds not only
in color-favored cases in which the current produces a light meson, where it
has been justified \cite{BBNS}, but also when the current produces a heavy
meson, where no such justification has been presented.  The importance of such
processes has recently been stressed by Lipkin \cite{HJL}.  
We also find that new experimental ratios of
helicity amplitudes for color-favored processes agree with predictions.
We shall ignore small non-factorizable contributions to color-favored $\ob$
decays as discussed, for example, in Ref.\ \cite{Terasaki}.

In Section II we review factorization predictions for the decays $\ob \to
W^{*-} D^{(*)+}$, where the virtual $W^{*-}$ produces either a lepton pair
$l^- \bar \nu_l$ or a hadron $\pi^-$, $\rho^-$, $a_1^-$, $D_s^-$,
or $D_s^{*-}$.  These processes are purely
color-favored.  We do not consider the corresponding $B^-$
decays, for which the nonleptonic processes receive both color-favored and
color-suppressed contributions.  We then (Section III) discuss the differential
distributions $d \Gamma(\ob \to D^{(*)+} l^- \bar \nu_l)/dq^2$ and the
information they can provide regarding the values of $|V_{cb}|$ and the form
factor slope at the normalization point.  Results of fits to $\ob$ two-body
decays to charmed final states 
are presented and compared with those from semileptonic decays in
Section IV.  We discuss the predictions of the factorization hypothesis for
decays in which the weak current produces a $D_s^{(*)}$ in Section V and for
ratios of helicity amplitudes in Section VI.  Section VII concludes.
An Appendix summarizes parameters of error ellipses used in combining 
data.

\section{Notation and predictions}

\begin{figure}
\centerline{\epsfysize = 2in \epsffile{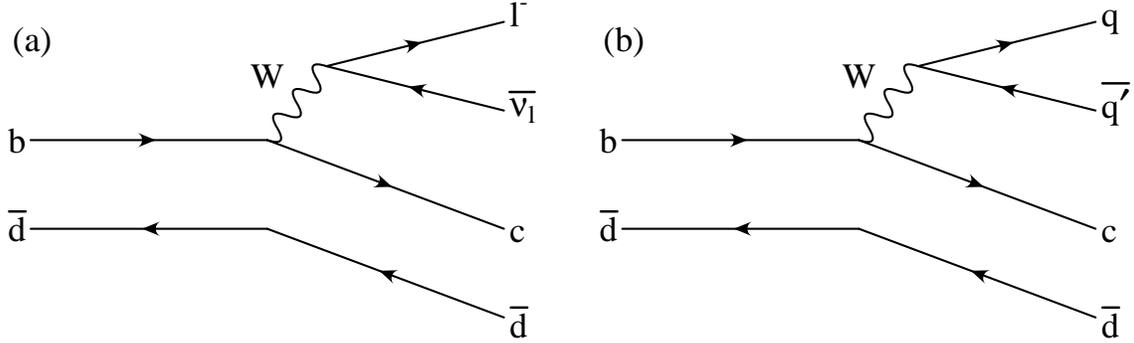}}
\caption{Feynman diagrams for $\ob$ decay illustrating the factorization
hypothesis.  A semileptonic decay (a) is related to a hadronic decay (b)
by the replacement of the lepton pair of 4-momentum $q$ with a $q \bar q'$
pair of effective mass $q^2$.
\label{fig:trees}}
\end{figure}

We review notation which is described in more detail in Ref.\ \cite{JRFM}.
We consider processes in which a semileptonic $\ob$ decay of the form shown in
Fig.\ \ref{fig:trees}(a) can be related to the corresponding decay with
the lepton pair replaced by a quark pair, illustrated in Fig.\
\ref{fig:trees}(b).  The matrix element for production of a pseudoscalar
meson $P(q)$ of 4-momentum $q$ from the vacuum by the axial vector current is
\beq
\bra{P(q)} A_\mu \ket{0} = i f_P q_\mu~~~,
\eeq
while that for production of a vector meson by the vector current is
\beq
\bra{V(q)} V_\mu \ket{0} = \epsilon^*_\mu M_V f_V~~~,
\eeq 
and that for production of an axial vector meson by the axial vector current is
\beq
\bra{A(q)} A_\mu \ket{0} = \epsilon^*_\mu M_A f_A~~~,
\eeq

% This is Figure 1

The form factors for the $\ob(v) \to D^{(*)}(v')$ transitions are described in
the heavy-quark limit by one universal function of the Lorentz-invariant
variable $w \equiv v \cdot v'$, where $v$ and $v'$ are invariant
four-velocities:  $v \equiv p_{\ob}/m_B$, $v' \equiv p_{D^{(*)}}/m_{D^{(*)}}$.
We take $c=1$ and note that $q = p - p'$.  Another expression for $w$ is then
\beq
w = \frac{p_B \cdot p_{D^{(*)}}}{m_B m_{D^{(*)}}} = \frac{m_B^2 + m^2_{D^{(*)}}
 - q^2} {2 m_B m_{D^{(*)}}}~~~.
\eeq
[A variable $(v - v')^2 = 2(1-w)$ was called $w$ in Ref.\ \cite{JRFM}.]
If $\epsilon$ denotes the polarization
vector of the final $D^*$, we may write \cite{BJ}
\beq
\bra{D(v')} V_\mu \ket{B(v)} = \sqrt{m_D m_B} F_V(w) (v + v')_\mu~~~,
\eeq
\beq
\bra{D^*(v',\epsilon)} A_\mu \ket{B(v)} = \sqrt{m_D m_B} F_A(w)
[\epsilon^*_\mu (1 + v \cdot v') - \epsilon^* \cdot v~v'_\mu]~~~,
\eeq
\beq
\bra{D^*(v',\epsilon)} V_\mu \ket{B(v)} = -i \sqrt{m_D m_B} F_V(w)
\epsilon_{\mu \nu \alpha \beta} \epsilon^{* \nu} v^\alpha {v'}^\beta~~~.
\eeq
We take $m_B = 5.28$ GeV and $m_D = m_{D^{*+}}=2.01$ GeV or $m_{D^+}
= 1.87$ GeV depending on the final-state charmed meson.
The maximum momentum transfer occurs when the recoiling $D^{(*)}$ is at rest in
the $B$ rest frame, so $q^2_{\rm max} = (m_B - m_{D^{(*)}})^2$
and hence $w \ge 1$.  Another useful relation is
\beq
y \equiv \frac{q^2}{m_B^2} = 1 - 2 w \sqrt{\zeta^{(*)}} + \zeta^{(*)}~~,
~~~\zeta^{(*)} \equiv \frac{m^2_{D^{(*)}}}{m_B^2}~~~.
\eeq

The differential decay width as a function of $w$ for
\mbox{$\ob \to D^{(*)+} l^- \bar{\nu_l}$} can then be written as
\beq \label{eqn:diff}
\frac{d\Gamma}{dw} = \frac{G^2_F}{48 \pi^3}|V_{cb}|^2 m_B^2 m_{D^{(*)}}^3
\sqrt{w^2 - 1} f^{(*)}(w) |F_{V,A}(w)|^2~~~,
\end{equation}
where for \mbox{$B \to D l \bar{\nu_l}$}
\begin{displaymath}
f(w) = (w^2-1)(1+\sqrt\zeta)^2~~~,
\end{displaymath}
for \mbox{$B \to D_T^{*} l \bar{\nu_l}$} 
\begin{displaymath}
f_T^*(w) = 4wy(w+1)~~~,
\end{displaymath}
and for \mbox{$B \to D_L^{*} l \bar{\nu_l}$} 
\begin{displaymath}
f_L^*(w) = (1 - \sqrt{\zeta^*})^2(w+1)^2~~~.
\end{displaymath}

We shall take the form factors $F_{V,A}(w)$ to be parametrized as in
Ref.~\cite{CLN}.  The form factor $F_V(w)$ governing the process $\ob \to D^+
l^- \bar{\nu}_l$ can be expressed as
\beq \label{eqn:CLNV}
F_V(w) = F_V(1) \times \left[ 1 - 8 \rfvs
z + (51 \rfvs - 10)z^2 - (252 \rfvs - 84)z^3 \right] ,
\eeq
where $z \equiv \mbox{$(\sqrt{w+1}-\sqrt{2})/(\sqrt{w+1}+\sqrt{2})$}$
(the corresponding variable for elastic $B \to B$ transitions, a natural one
for discussing analyticity in dispersion relations, was introduced
in \cite{BGL95}),
while the form factor $F_A(w)$ governing $B\to D^* l \bar{\nu}_l$ is
related to the axial-vector form factor $A_1(w)$ by
$$
\left[ 1 + \frac{4w}{w+1} \frac{1-2w\sqrt{\zeta^*}+\zeta^*} 
{(1-\sqrt{\zeta^*})^2} \right] \left| F_A(w) \right| ^2 =
$$
\beq \label{eqn:CLNA}
\left\{ 2 \frac{1-2w\sqrt{\zeta^*}+\zeta^*}{(1- \sqrt{\zeta^*})^2}
\left[ 1 + \frac{w-1}{w+1} R_1(w)^2 \right]
 + \left[ 1+\frac{w-1}{1-\sqrt{\zeta^*}} 
\left(1-R_2(w)\right) \right]^2 \right\} \left| A_1(w) \right|^2.
\eeq
$A_1(w)$ can similarly be parametrized as
\begin{equation}
A_1(w) = A_1(1) \times \left[ 1 - 8 \ras z + (53 \ras - 15)z^2
- (231 \ras - 91)z^3 \right]~~.
\end{equation}
These forms are motivated by dispersion relations \cite{CLN,BGL95,BGL}. 
$R_1(w)$ and $R_2(w)$ are given by
\begin{eqnarray}
R_1(w)&=&R_1(1)-0.12(w-1)+0.05(w-1)^2, \nonumber \\
R_2(w)&=&R_2(1)+0.11(w-1)-0.06(w-1)^2.
\end{eqnarray}
In this paper we use the CLEO experimental results for $R_1(1)$ and
$R_2(1)$ \cite{CLEOVcb}:
\begin{eqnarray}
R_1(1)&=&1.18\pm 0.30 \pm 0.12, \\
R_2(1)&=&0.71\pm 0.22 \pm 0.07.
\end{eqnarray}
As we know, $\rfvs$ and $\ras$ are the slope parameters for the
form factors $F_V(w)$ and $A_1(w)$ at zero recoil, respectively. The
difference between $\ras$ and $\rfas$, the slope parameter for
$F_A(w)$ at $w = 1$, is predicted to be $\rfas - \ras = 0.21$
\cite{CLN}. Recall that $\rfvs$ and $\rfas$ are actually the
same in the single pole model. To make a connection between
$F_{V,A}(w)$ and the single pole form factor \cite{JRFM}:
\begin{equation} \label{eqn:pole}
{\mathcal F}_{V,A}(w) = {\mathcal F}_{V,A}(1)/[1-2(1-w)/w^2_{0(V,A)}],
\end{equation}
we assume $\rfvs = \ras - 0.21$.  From now on we will simplify $\ras$ as $\rs$.
This parameter describes the slope of the Isgur-Wise \cite{IW} form factor
at the zero-recoil point:  $\rho^2 = [dF_{A_1}(w)/dw]|_{w=1}$ (see, e.g.,
\cite{COV}).

At $w=1$, the vector and axial vector form factors are expected to behave as
$F_V(1) = \eta_V(1 + \delta_{1/m_b})$, $F_A(1) = \eta_A(1 + \delta_{1/m_b^2})$.
Here $\eta_V = 1.022 \pm 0.004$ and $\eta_A = 0.960 \pm 0.007$ are QCD
corrections \cite{Cz}.  The terms $\delta_{1/m_b}$ and $\delta_{1/m_b^2}$ are
non-perturbative in origin, and correspond physically to the excitation of
states other than $D$ and $D^*$. Lacking a reliable method for estimating the
term $\delta_{1/m_b}$ in $F_V(1)$ \cite{BLRW}, we set it equal to zero.
The absence of a $\delta_{1/m_b}$ term in $F_A(1)$ is the subject of Luke's
theorem \cite{Luke}.  We take $\delta_{1/m_b^2} = -0.05 \pm 0.035$
\cite{Neub,FalkTASI}, with the product $F_A(1) = 0.913 \pm 0.042$ as used in
Ref.~\cite{CLEOVcb}.  Eq.\ (\ref{eqn:diff}) can then be integrated with respect
to $w$ to yield predicted decay rates as functions of the two parameters
$\rs$ and $|V_{cb}|$.

The decay widths of some nonleptonic modes may be obtained under the
assumption of factorization.  For simplicity
we assume all $\ob \to D^+ M^-$ transitions to involve $F_V(w_M)$ and all
$\ob \to D^{*+} M^-$ transitions to involve $F_A(w^*_M)$, where
\beq
w_M \equiv (m_B^2 + m_D^2 - m_M^2)/(2 m_B m_D)~~,~~~
w^*_M \equiv (m_B^2 + m_{D^*}^2 - m_M^2)/(2 m_B m_{D^*})~~~.
\eeq
We then find
\begin{eqnarray}
\Gamma(\ob \to D^+ \pi^-) & = & \frac{G_F^2}{32\pi} |V_{cb}|^2 |V_{ud}|^2
 m_B^3 f_{\pi}^2 |a_1|^2 |F_V(w_{\pi})|^2 (1-\sqrt\zeta)^2 \nonumber \\ 
& & \hspace{6ex} \times \lambda^{1/2}(1,\zeta, \zeta_{\pi})
 \frac{[(1+\sqrt\zeta)^2-\zeta_{\pi}]^2}{4\sqrt\zeta} \label{eqn:dp}\\
\Gamma(\ob \to {D^*}^+ \pi^-) &=&\frac{G_F^2}{32\pi} |V_{cb}|^2
 |V_{ud}|^2 m_B^3 f_{\pi}^2 |a_1|^2 |F_A(w_{\pi}^*)|^2 (1+\sqrt{\zeta^*})^2
 \nonumber \\
& & \hspace{6ex} \times \lambda^{1/2}(1,\zeta^*,\zeta_{\pi})
 \frac{\lambda(1,\zeta^*,\zeta_{\pi})}{4\sqrt{\zeta^*}} \label{eqn:dsp} \\
\Gamma(\ob \to {D}^+\rho^-) &=&\frac{G_F^2}{32\pi} |V_{cb}|^2 |V_{ud}|^2
 m_B^3 f_{\rho}^2 |a_1|^2 |F_V(w_{\rho})|^2 (1+\sqrt\zeta)^2 \nonumber \\
& & \hspace{6ex} \times \lambda^{1/2}(1,\zeta,\zeta_{\rho})
 \frac{\lambda(1,\zeta,\zeta_{\rho})}{4\sqrt\zeta} \label{eqn:dr} \\
\Gamma(\ob \to {D^*}^+\rho^-) &=&\frac{G_F^2}{32\pi}|V_{cb}|^2 |V_{ud}|^2
 m_B^3 f_{\rho}^2 |a_1|^2 |F_A(w_{\rho}^*)|^2 N(\zeta^*,\zeta_{\rho})
 \nonumber \\
& & \hspace{6ex} \times \lambda^{1/2}(1,\zeta^*,\zeta_{\rho})
 \frac{(1+\sqrt{\zeta^*})^2-\zeta_{\rho}}{4\sqrt{\zeta^*}} \label{eqn:dsr} \\
\Gamma(\ob \to D^+ a_1^-) &=&\frac{G_F^2}{32\pi} |V_{cb}|^2 |V_{ud}|^2
 m_B^3 f_{a_1}^2 |a_1|^2 |F_V(w_{a_1})|^2 (1+\sqrt\zeta)^2 \nonumber \\
& & \hspace{6ex} \times \lambda^{1/2}(1,\zeta,\zeta_{a_1})
 \frac{\lambda(1,\zeta,\zeta_{a_1})}{4\sqrt\zeta} \label{eqn:da} \\
\Gamma(\ob \to {D^*}^+a_1^-) &=&\frac{G_F^2}{32\pi}|V_{cb}|^2 |V_{ud}|^2
 m_B^3 f_{a_1}^2 |a_1|^2 |F_A(w_{a_1}^*)|^2 N(\zeta^*,\zeta_{a_1}) \nonumber \\
& & \hspace{6ex} \times \lambda^{1/2}(1,\zeta^*,\zeta_{a_1})
 \frac{(1+\sqrt{\zeta^*})^2-\zeta_{a_1}}{4\sqrt{\zeta^*}} \label{eqn:dsa}
\end{eqnarray}
where $\zeta_M = m_M^2/m_B^2$ while
\begin{equation}
N(\zeta^*,\zeta_M)\equiv (1-\sqrt{\zeta^*})^2[(1+\sqrt{\zeta^*})^2
-\zeta_M]+4\zeta_M(1+\zeta^*-\zeta_M)~~,
\end{equation}
\beq
\lambda(a,b,c) \equiv a^2 + b^2 + c^2 - 2ab - 2ac - 2bc~~.
\eeq
The QCD correction $|a_1|$ is taken to be 1.05 for all processes; this is
a sufficiently good approximation to the actual situation, in which values
differ by less than a percent from process to process \cite{BBNS}.  In the
limit of small $m_\pi$, the results (\ref{eqn:dp}) and (\ref{eqn:dsp})
are special cases of the simple Bjorken relation \cite{BJ}
\beq
\Gamma(\ob \to D^{(*)+} \pi^-) = 6 \pi^2 f_\pi^2 |V_{ud}|^2 |a_1|^2 \left.
\frac{d \Gamma(\ob \to D^{(*)+} l^- \bar \nu_l)}{d q^2} \right|_{q^2 = m_\pi^2}
~~~.
\eeq

\section{Semileptonic decays}

\begin{figure}
\centerline{\epsfysize = 5.5in \epsffile{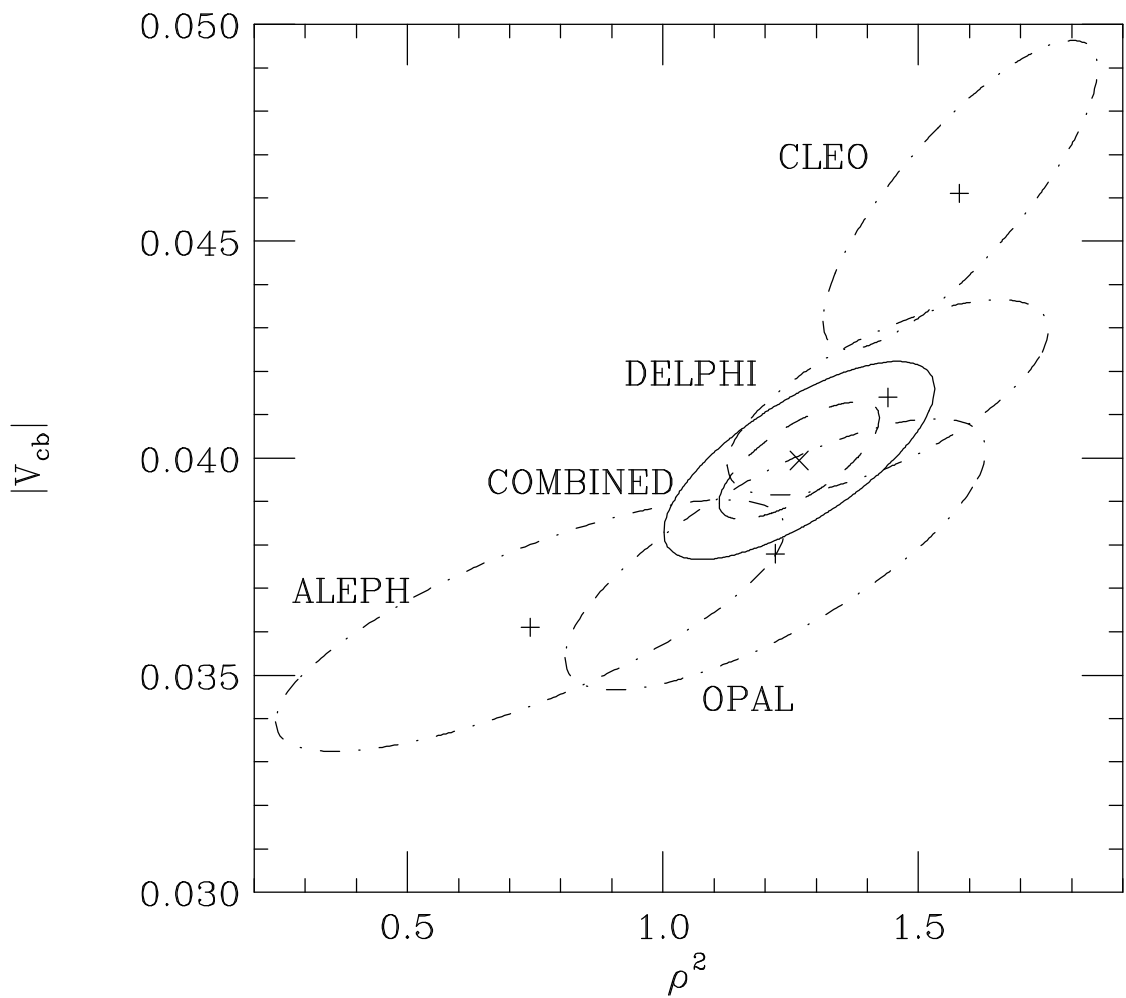}}
\caption{Error ellipses corresponding to $\Delta \chi^2 = 1$ for fits to
$\ob \to D^{*+} l^- \bar \nu_l$ spectra. Dot-dashed lines correspond to fits to
individual experiments, not including theoretical error in $F_A(1)$.  Fit to
combined semileptonic data without (with) error in $F_A(1)$ is shown by the
dashed (solid) ellipse.  The errors we quote on each variable in Table
\ref{tab:slep}
correspond to $\pm 1 \sigma$ extremes in which the other variable is also
permitted to vary.  The plotted points ($+$ for individual experiments,
$\times$ for combined data) correspond to $\chi^2$ minima.
\label{fig:slep}}
\end{figure}

The CLEO Collaboration \cite{CLEOVcb} at the Cornell Electron Storage
Ring (CESR) and the ALEPH, DELPHI, and OPAL Collaborations
\cite{ALEPH,DELPHI,OPAL} at LEP \cite{LEP} have measured the spectra
in lepton-pair squared effective mass $q^2$ (equivalently, in the Isgur-Wise
variable $w$) for the decay $\ob \to D^{*+} l^- \bar \nu_l$.  The spectra
may then be fitted for $F_A(1)|V_{cb}|$ and $\rs$.  There is a strong
correlation between the two parameters.  The results are shown in
Fig.~\ref{fig:slep} and Table \ref{tab:slep}, where we have taken $F_A(1) =
0.913 \pm 0.042$ as in Ref.\ \cite{CLEOVcb}. Our fitted parameters for
the CLEO data differ slightly from those presented by the collaboration itself,
since we wished to generate an error ellipse and therefore fitted the
spectral points directly without taking account of point-to-point correlations.
For comparison, CLEO quotes $|V_{cb}| = 0.0464 \pm 0.0020({\rm stat.})
\pm 0.0021({\rm syst.}) \pm 0.0021({\rm theor.})$ and $\rs = 1.67 \pm 0.11
\pm 0.22$.  The combined fit implies $|V_{cb}| = 0.0399 \pm 0.0023$ and
$\rs = 1.27 \pm 0.26$.

% This is Figure 2

% This is Table I
\begin{table}
\caption{Values of $|V_{cb}|$ and $\rs$ obtained from fits to individual
spectra in $\ob \to D^{*+} l^- \bar \nu_l$ decays. \label{tab:slep}}
\begin{center}
\protect
\begin{tabular}{c c c c} \hline
Experiment & $|V_{cb}|$ & $\rs$ & \\ \hline
CLEO \cite{CLEOVcb}  & $0.0461 \pm 0.0036$ & $1.58 \pm 0.27$ & \\
ALEPH \cite{ALEPH}   & $0.0361 \pm 0.0029$ & $0.74 \pm 0.50$ & \\
DELPHI \cite{DELPHI} & $0.0378 \pm 0.0031$ & $1.22 \pm 0.41$ & \\
OPAL \cite{OPAL}     & $0.0414 \pm 0.0023$ & $1.44 \pm 0.32$ & \\
Combined (a)         & $0.0399 \pm 0.0023$ & $1.27 \pm 0.26$ & \\ \hline
\end{tabular}
\end{center}
\leftline{\qquad \qquad (a) Errors include common theoretical error on
 $F_A(1)$.}
\end{table}

The CLEO Collaboration has also measured the spectrum for the decay
$\ob \to D^+ l^- \bar \nu_l$ \cite{CLEOBD}.  A fit to this spectrum with the
form factor (10) and with $F_V(1) = 1.022$ yields $|V_{cb}| =
0.0459^{+0.0053}_{-0.0044}$ and $\rho^2 = 1.33^{+0.21}_{-0.25}$.

\section{Nonleptonic two-body decays and combined fit}

% This is Table II
\begin{table}
\caption{Branching ratios for $\ob$ decays averaged for our fits, in units
of $10^{-3}$.  \label{tab:avgs}}
\begin{center}
\protect
\begin{tabular}{c c c c} \hline
Mode           & Value & Value & Average \\ \hline
$D^{*+} \pi^-$ & $2.76 \pm 0.21$ \cite{PDG} & $2.9 \pm 0.3 \pm 0.3$
 \cite{BaD} & $2.79 \pm 0.19$ \\
$D^{*+} \rho^-$ & $6.8 \pm 3.4$ \cite{PDG} & $11.2 \pm 1.1 \pm 2.5$
 \cite{BaD} & $9.5 \pm 2.1$ \\
$D^{*+} D_s^-$ & $11.0 \pm 1.8 \pm 1.0 \pm 2.8$ \cite{ClDs} &
 $7.1 \pm 2.4 \pm 2.5 \pm 1.8$ \cite{BaDs} & $10.0 \pm 3.1$ \\
$D^{*+} D_s^{*-}$ & $18.2 \pm 3.7 \pm 2.4 \pm 4.6$ \cite{ClDs} &
 $25.4 \pm 3.8 \pm 5.3 \pm 6.4$ \cite{BaDs} & $20.5 \pm 6.3$ \\ \hline
\end{tabular}
\end{center}
\end{table}

We fit rates for nonleptonic two-body decays [Eqs.\
(\ref{eqn:dp}--\ref{eqn:dsa})] to experimental averages, allowing (as in
the fit to semileptonic spectra) for variation of $|V_{cb}|$ and $\rs$.
We use the averages of Ref.\ \cite{PDG} except for modes for which new values
have been presented \cite{BaD,ClDs,BaDs}; these are summarized and averaged in
Table \ref{tab:avgs}.  For the decays involving $D^{(*)}_s$ (to be discussed in
the next Section) the last errors in the 
second and third columns of Table \ref{tab:avgs} refer to a common systematic
error in $D_s$ branching ratios, which are based on ${\cal B}(D^+_s \to \phi
\pi^+) = (3.6 \pm 0.9)\%$, and are combined accordingly.  In our fits we use
$|V_{ud}| = 0.974$, $\tau_B = 1.548\hspace{1ex}{\rm ps}$, and
$f_{\pi} = 131\hspace{1ex}{\rm MeV}$. $f_{\rho}$ and $f_{a_1}$ are determined
to be $209\hspace{1ex}{\rm MeV}$ and $229 \hspace{1ex}{\rm MeV}$
\cite{BBNS}, respectively,
from the branching ratios for $\tau \to \rho \nu$ and $\tau \to a_1 \nu$.

\begin{figure}
\centerline{\epsfysize = 5.5in \epsffile{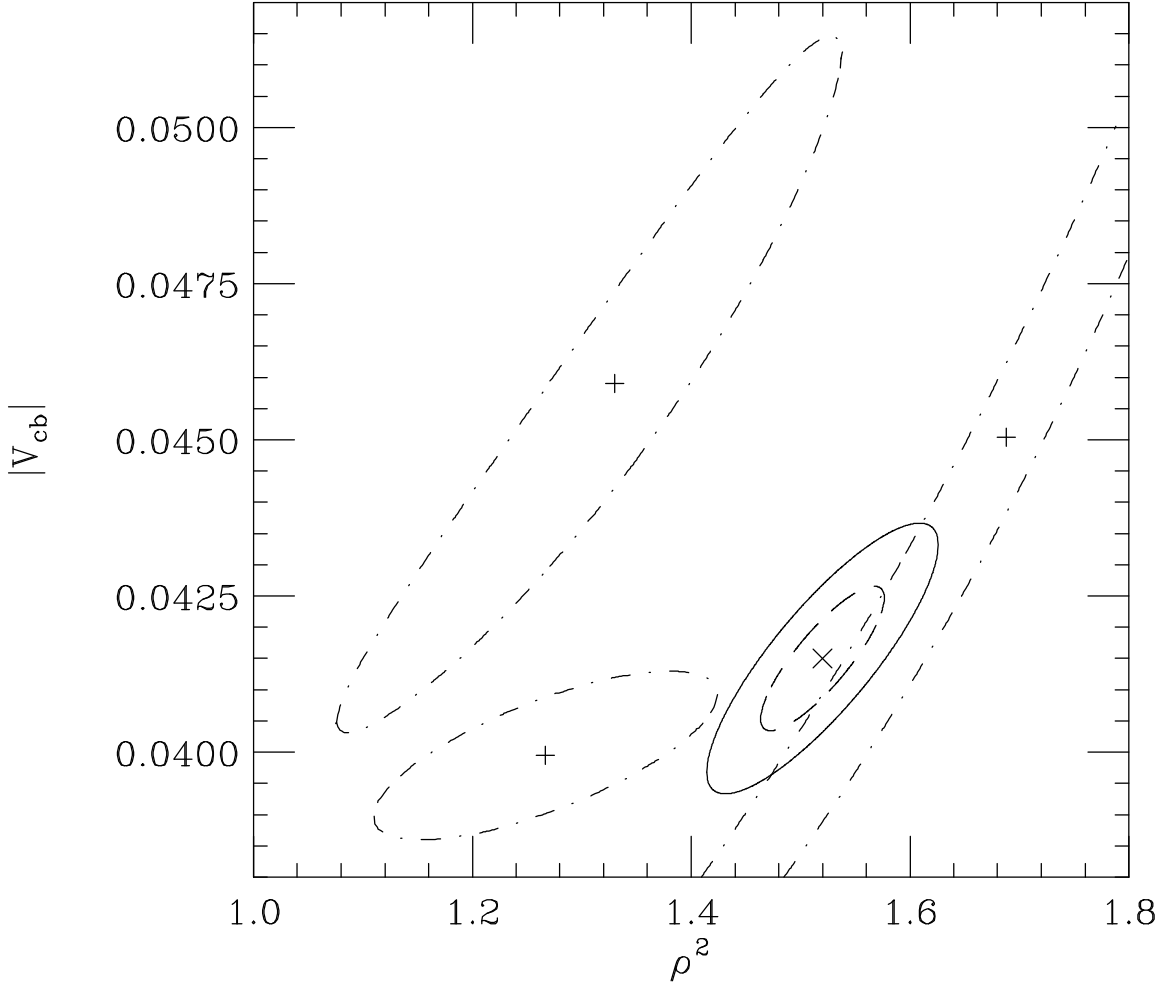}}
\caption{Right-hand pair of dash-dotted curves: Contours of $\Delta
\chi^2 = 1~(1 \sigma)$ for fit to nonleptonic two-body
decays $\ob \to D^{(*)+} (\pi^-,\rho^-,a_1^-)$.  Upper dash-dotted ellipse:
Contour of $\Delta \chi^2 = 1~(1 \sigma)$ for fit to $\ob \to D^+ l^- \bar
\nu_l$ spectrum.  Lower dash-dotted ellipse: 
Combined fit to semileptonic $\ob \to D^{*+} l^- \bar \nu_l$ decays.  All
these fits are performed assuming $F_A(1) = 0.913$ and $F_V(1) = 1.022$.
Contour of $\Delta \chi^2 = 1~(1 \sigma)$ for fit to combined semileptonic and
nonleptonic data without (with) common theoretical error in form factor
normalization is shown by the dashed (solid) ellipse.  $\chi^2$ minima are
indicated by $+$ for nonleptonic and $\ob \to
D^{(*)+} l^- \bar \nu_l$ decays and by $\times$ for combination of all data.
\label{fig:comb}}
\end{figure}

The fit to two-body nonleptonic decays alone gives rise to a different
correlation between $|V_{cb}|$ and $\rs$ than that to the $\ob \to D^{(*)+}
l^- \bar \nu_l$ spectra, since the decay rates are dominated by low $q^2$ and
hence high $w$.  Contours of $\Delta \chi^2 = 1~(1 \sigma$) for nonleptonic
decays are shown along with the $\Delta \chi^2 = 1$ contours for
$\ob \to D^{(*)+} l^- \bar \nu_l$ spectra in Fig.\ 
\ref{fig:comb}.  Also shown are the $\Delta \chi^2 = 1$ contours for the
combined fit without and with common theoretical errors.  We find $|V_{cb}|
= 0.0415 \pm 0.0022$ and $\rs = 1.52
\pm 0.11$.  The results are summarized in Table \ref{tab:comb}.  The error
on $|V_{cb}|$ is dominated by the theoretical uncertainty on the form factors
at $w=1$, which we take to have the same fractional value (0.042/0.913) for
vector and axial form factors. 

% This is Table III
\begin{table}
\caption{Values of $|V_{cb}|$ and $\rs$ obtained from fits to nonleptonic
two-body $\ob \to D^{(*)+} (\pi^-,\rho^-,a_1^-)$ decays and
$\ob \to D^{(*)+} l^- \bar \nu_l$ decays. \label{tab:comb}}
\begin{center}
\protect
\begin{tabular}{c c c} \hline
Decays        &    $|V_{cb}|$         & $\rs$        \\ \hline
Nonleptonic   &    $0.0450$ (a) & $1.69$ (a) \\
$\ob \to D^{*+} l^- \bar \nu_l$ spectra & $0.0399 \pm 0.0023$ & $1.27 \pm 0.26$ \\
$\ob \to D^+ l^- \bar \nu_l$ spectrum & $0.0459^{+0.0053}_{-0.0044}$ (a)
& $1.33^{+0.21}_{-0.25}$ (a) \\
Combined (b)  & $0.0415 \pm 0.0022$ & $1.52 \pm 0.11$ \\ \hline
\end{tabular}
\end{center}
\leftline{\qquad (a) Large correlated errors; see Fig.\ \ref{fig:comb}.}
\leftline{\qquad (b) Errors include common theoretical error of $\delta F_A(1)/
F_A(1) = 4.6\%$.}
\end{table}
 
% This is Figure 3

% This is Table IV
\begin{table}[!htb]
\caption{Branching ratios in units of $10^{-3}$: comparison between data and
predictions.
\label{tab:fit}}
\begin{center}
\begin{tabular}{ l c c c c }
\hline
Decay mode & Data & Ref.\ \cite{BBNS} (a) & Our fit & $\chi^2$ contribution
\\ \hline
$\ob \to D^+ \pi^-$ & $3.0 \pm 0.4$ & 3.27 & 3.19 & 0.22 \\
$\ob \to D^{*+} \pi^- $ & $2.79 \pm 0.19$ & 3.05 & 3.10 & 2.70 \\
$\ob \to D^+ \rho^-$ & $7.9 \pm 1.4$ & 7.64 & 7.92 & 0.00 \\
$\ob \to D^{*+} \rho^-$ & $9.5 \pm 2.1$ & 7.59 & 8.78 & 0.12 \\
$\ob \to D^+ a_1^-$ & $6.0 \pm 3.3$ & 7.76 & 9.10 & 0.88 \\
$\ob \to D^{*+} a_1^-$ & $13.0 \pm 2.7$ & 8.53 & 12.2 & 0.09 \\ 
\hline
\end{tabular}
\end{center}
\leftline{\qquad \qquad (a) For preferred values of form factors and $|a_1| =
1.05$.}
\end{table}

The fitted branching ratios are compared with experimental data in Table 
\ref{tab:fit}.  We also show the predictions
of a recent investigation based on a more detailed application of the
factorization hypothesis \cite{BBNS}.  The quality of the fit is acceptable
except for a slight excess in the predicted branching ratio for $\ob \to 
D^{*+} \pi^-$.

\begin{figure}
\centerline{\epsfysize = 5.5in \epsffile{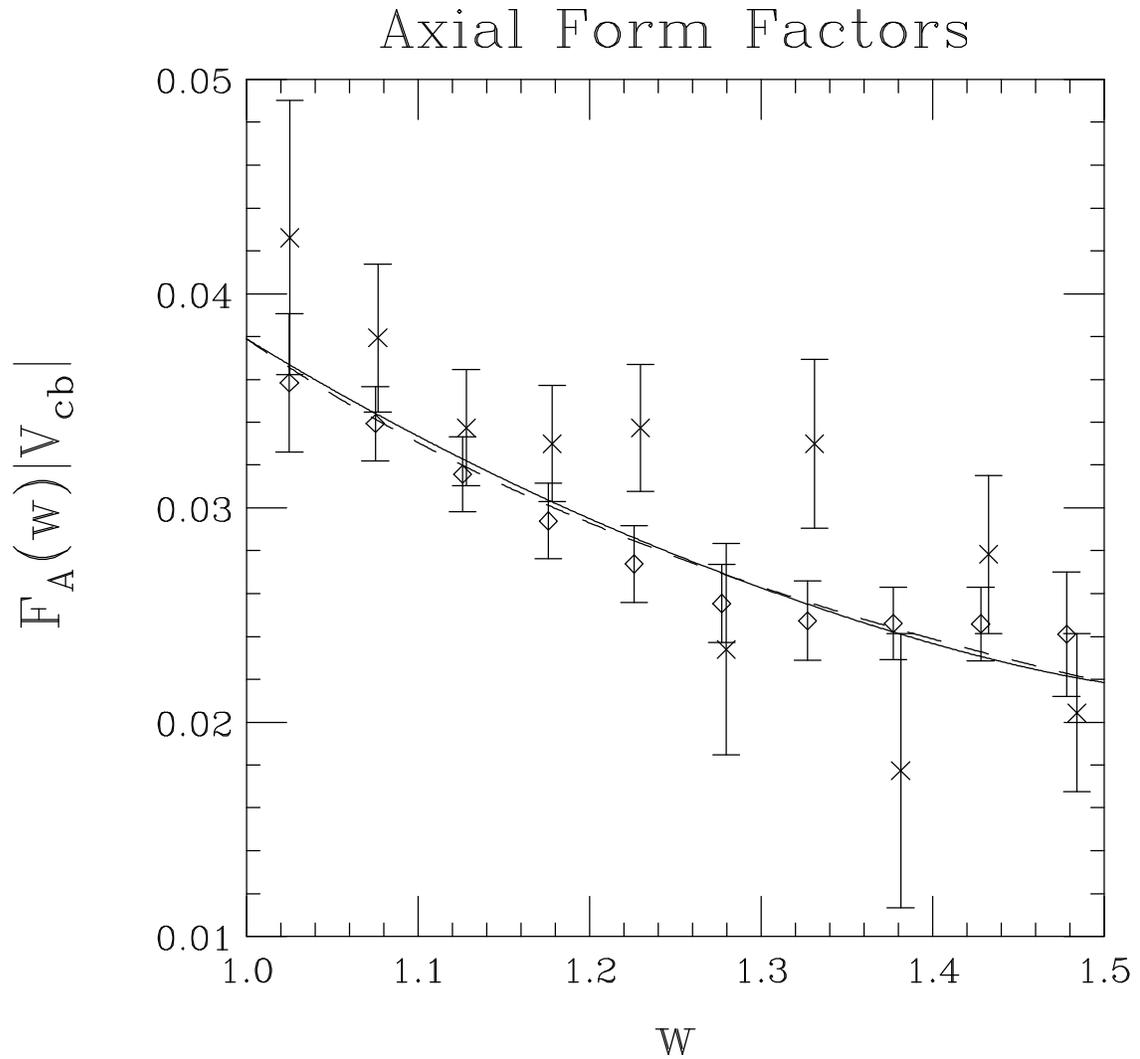}}
\caption{Form factors (\ref{eqn:CLNA}) (solid curve) for $\ras = 1.52$
and pole-model (\ref{eqn:pole}) (dashed curve) for $w_{0A} = 1.17$.  Data
points are from CLEO (crosses) and DELPHI (diamonds). \label{fig:afmfs}}
\end{figure}

It is interesting to compare the form factors based on Eqs.~(\ref{eqn:CLNV})
and (\ref{eqn:CLNA}) with the simple pole model (\ref{eqn:pole}) \cite{JRFM},
where $m_B w_{0(V,A)}$ has the interpretation of the mass of a pole in the
weak $b \to c$ (V,A) current.  The axial form factor for $\ras = 1.52$ is
compared with a pole model with $w_{0A} = 1.17$ in Fig.~\ref{fig:afmfs}.  Also
shown are CLEO \cite{CLEOVcb} and DELPHI \cite{DELPHI} data points.  The
pole-model form factor is almost indistinguishable from that \cite{CLN}
motivated by dispersion relations.  The value $w_{0A} = 1.12 \pm 0.17$,
found in Ref.~\cite{JRFM}, is consistent with the present determination
$w_{0A} = 1.17 \pm 0.08$.

\begin{figure}
\centerline{\epsfysize = 5.5in \epsffile{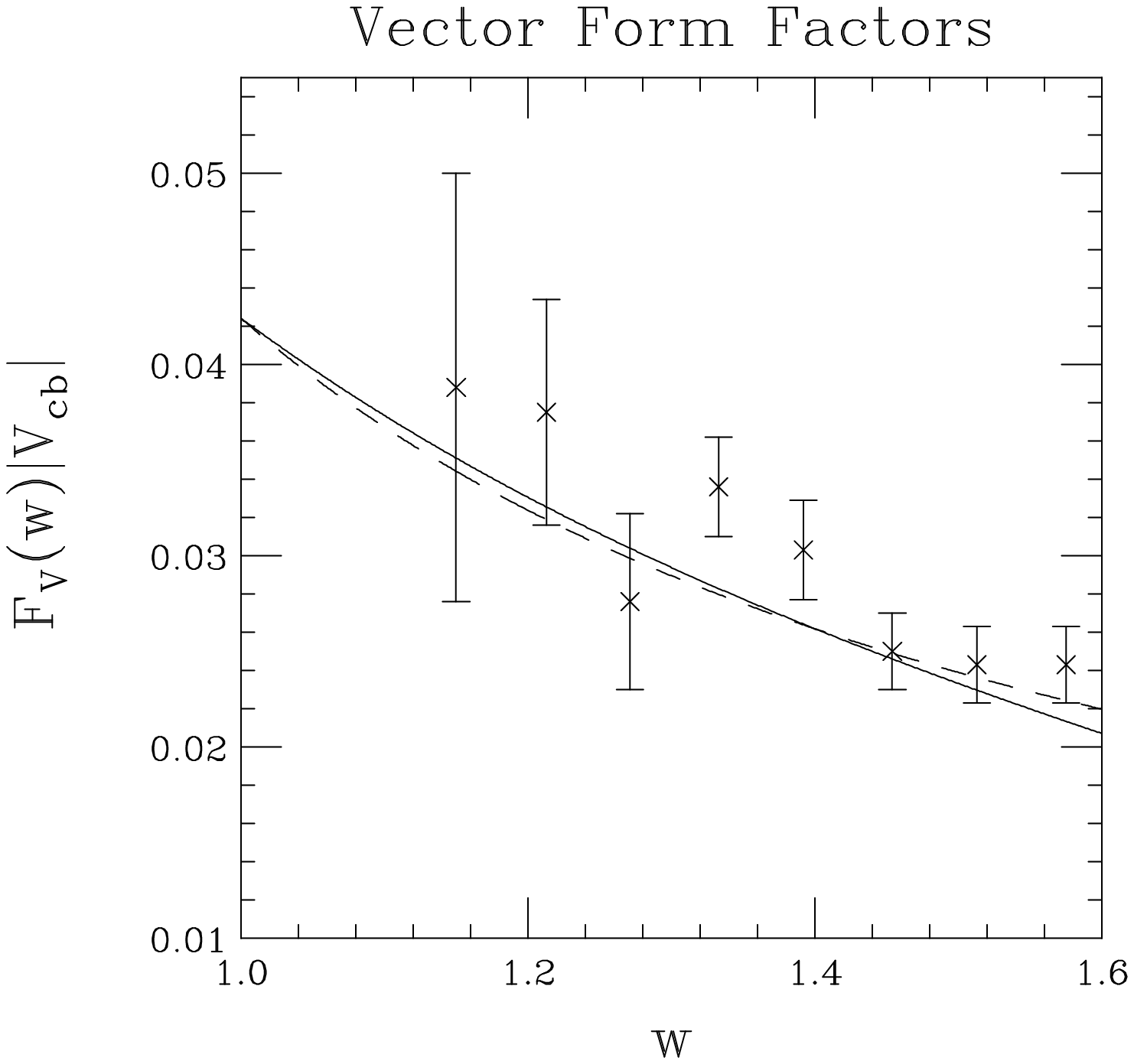}}
\caption{Form factors (\ref{eqn:CLNV}) (solid curve) for $\rfvs = 1.31$
and pole-model (\ref{eqn:pole}) (dashed curve) for $w_{0V} = 1.14$ compared
with CLEO data (plotted points).
\label{fig:vfmfs}}
\end{figure}

The vector form factor (\ref{eqn:CLNV}) is characterized by a slope
parameter $\rfvs = \ras - 0.21$ \cite{CLN} and hence $\rfvs =
1.31$.  It is compared to a pole form factor with $w_{0V} = 1.14$ and
to the CLEO data \cite{CLEOBD} in
Fig.~\ref{fig:vfmfs}.  Thus, a nearly universal pole position characterizes
the vector and axial form factors, as in Ref.\ \cite{JRFM}.  The CLEO data
lie slightly above the predicted form factor but have the same $w$ dependence,
as one can also see in Fig.\ 3.  It should be recalled that, in contrast to the
case of the axial-vector form factor, theoretical estimates of the ${\cal O}
(1/m_b)$ correction to the vector form factor are lacking \cite{BLRW}.  The
normalization of the CLEO data may reflect our ignorance of this correction.

\section{$D^{(*)}_s$ production by the weak current}

When the $q \bar q'$ meson in Fig.\ \ref{fig:trees}(b) is a $D^-_s$ or
$D_s^{*-}$, Eqs.\ (\ref{eqn:dp}--\ref{eqn:dsa}) can be used for the
respective predictions for $\Gamma(\ob \to D^+ D_s^-)$, $\Gamma(\ob \to
D^{*+} D_s^-)$,  $\Gamma(\ob \to D^+ D_s^{*-})$, and $\Gamma(\ob \to
D^{*+} D_s^{*-})$ by replacing $f_\pi \to f_{D_s}$, $f_\rho \to f_{D^*_s}$,
$\zeta_\pi \to \zeta_{D_s} \equiv m^2_{D_s}/m_B^2$, $\zeta_\rho \to
\zeta_{D^*_s} \equiv m^2_{D^*_s}/m_B^2$, and other corresponding substitutions
of kinematic variables $w$ and $N$.  We shall assume $f_{D^*_s} = f_{D_s} = 270
\pm 16 \pm 34$ MeV based on an experimental average of rates for $D_s \to \mu
\nu$ and $D_s \to \tau \nu$ \cite{JJT}; the latter error is the common
systematic error associated with a 25\% uncertainty in the branching ratio
for $D_s^+ \to \phi \pi^+$.
(For comparison, the value of $259 \pm 74$ MeV was found in Ref.\ \cite{JRFM}
by utilizing the observed branching ratios of processes in which the weak
current produced a $D_s$ or $D_s^*$.)

We then predict the branching ratios for $D_s^{(*)}$ production by the weak
current shown in Table \ref{tab:ds}.  Experimental values are from Ref.\
\cite{PDG} ($D^+ D_s^{(*)-}$) or Table \ref{tab:avgs} ($D^{*+} D_s^{(*)-}$).
The predictions have an overall 25\% uncertainty associated with all
$D_s$ branching ratios.  They are as well obeyed as those for the light mesons.

% This is Table V
\begin{table}
\caption{Comparison of predictions for branching ratios (in units of $10^{-3}$)
involving
$D_s^{(*)}$ production by the weak current in $\ob$ decays.  \label{tab:ds}}
\begin{center}
\begin{tabular}{l c c}  \hline
Decay mode & Data & Prediction \\ \hline
$D^+ D_s^-$ & $8.0 \pm 3.0$ & $14.9 \pm 4.1$ \\
$D^{*+} D_s^-$ & $10.0 \pm 3.1$ & $8.6 \pm 2.4$ \\
$D^+ D_s^{*-}$ & $10.0 \pm 5.0$ & $10.0 \pm 2.8$ \\
$D^{*+} D_s^{*-}$ & $20.5 \pm 6.3$ & $24.0 \pm 6.7$ \\ \hline
\end{tabular}
\end{center}
\end{table}

An additional prediction involving heavy meson production by the weak current
\cite{JRFM} is that ${\cal B}(\ob \to D^{*+} D^{*-})/{\cal B}(\ob \to D^{*+}
D_s^-) = 0.13(f_D/f_{D_s})^2 \simeq 0.09$, where $f_D$ and $f_{D_s}$ are
the decay constants for the nonstrange and strange $D$ mesons.  The
experimental value for this ratio \cite{CLEODD} is $0.06^{+0.04}_{-0.03}$.

\section{Ratios of helicity amplitudes}

The decays of spinless particles to two
vector mesons are describable \cite{DDLR} by amplitudes
$A_0$ (longitudinal polarization), $A_\parallel$ (linear parallel polarization)
and $A_\perp$ (linear perpendicular polarization), normalized such that
$|A_0|^2 + |A_\parallel|^2 + |A_\perp|^2 = 1$.  Factorization predicts
$(|A_0|^2, |A_\parallel|^2, |A_\perp|^2) = (88,10,2)\%$ for $\ob \to D^{*+}
D_s^{*-}$ and (55,39,6)\% for $\ob \to D^{*+} \rho^-$.  Experimental values
are only quoted for $|A_0|^2$: $(87.8 \pm 3.4 \pm 3.0)\%$ for $\ob \to D^{*+}
D_s^{*-}$ \cite{CLEODD} and $(50.6 \pm 13.9 \pm 3.6)\%$ for $\ob \to D^{*+}
\rho^-$ \cite{dr}.  These agree with the predictions, as does the
intermediate case of $\rho'(1418)$ production \cite{rhop}.

\section{Conclusions}

New data on $B$ meson decays have improved the precision of tests of some
early factorization predictions \cite{JRFM}, and yield a value $|V_{cb}| =
0.0415 \pm 0.0022$ when CLEO and LEP data on $\ob \to D^{(*)+} l^- \bar \nu_l$
spectra are combined with two-body nonleptonic decays $\ob \to D^{(*)+}
(\pi^-,\rho^-,a_1^-)$.
The slope of the universal Isgur-Wise form factor at the normalization
point $w=1$ is found to be described by the parameter $\rs = 1.52 \pm 0.11$.
These values are only slightly different from those based on $\ob \to D^{(*)+} 
l^- \bar \nu_l$ spectra alone, indicating that factorization for color-favored
decays and universality of $\overline{B} \to D^{(*)}$ form factors are
reasonable approximations.  Consistency between nonleptonic and semileptonic
determinations is at least as good as that among the semileptonic processes
themselves.  Our neglect of ${\cal O}(1/m_b)$ corrections to the vector form
factor may underestimate the rate for $\ob \to D^+ l^- \bar \nu_l$ slightly.

Satisfactory rates for processes involving $D_s^{(*)}$ production by the weak
current are obtained when the world average of direct measurements for
$f_{D_s}$ is used, and when it is assumed that $f_{D^*_s} = f_{D_s}$.
Ratios of helicity amplitudes for color-favored
processes are also found to be in accord with predictions.

\section*{Acknowledgments}

We thank Elisabetta Barberio, Karl Ecklund, Roger Forty, Richard Hawkings,
Jon Thaler, and Alan Weinstein for discussions.
This work was supported in part by the United
States Department of Energy through Grant No.\ DE FG02 90ER40560, and in part
by the U. S. -- Israel Binational Science Foundation through Grant No.\
98-00237.

\section*{Appendix:  Parametrization of error ellipses}

In order to combine results of fits to semileptonic decays in the absence of
the ALEPH and DELPHI raw spectra, we have parametrized their fits in terms
of error ellipses corresponding to $\Delta \chi^2 = 1$, and generated
corresponding ellipses for our own fits to CLEO and DELPHI data.  The
equations describing these ellipses are given below.  Also shown are
the contributions to $\Delta \chi^2$ for each set of $\ob \to D^{*+} l^-
\bar \nu_l$ data
when $x \equiv \rs$ and $y \equiv |V_{cb}|$ are taken to equal their
values $x = 1.52$ and $y = 0.0415$ in the global fit.  The corresponding
$\Delta \chi^2$ value is $4.02$ for the sum of the nonleptonic modes
listed in Table IV, which, when combined with the $\Delta \chi^2$ values for
the semileptonic spectra, leads to a total of $\Delta \chi^2 = 23.2$ for the
fit to five semileptonic spectra and six nonleptonic decay rates.  The largest
source of this $\Delta \chi^2$ (10.3) is the higher overall scale of the $\ob
\to D^+ l^- \bar \nu_l$ spectrum measured by CLEO, with some contribution also
from the disparity between the CLEO and ALEPH fits.  In comparison with these
disagreements among purely semileptonic processes, the fits to nonleptonic
decays do not fare badly at all.
\bigskip

CLEO:
\begin{eqnarray*}
41.8814(x-x_c)^2-5174.3(x-x_c)(y-y_c)+238793(y-y_c)^2&=&1;\\
x_c=1.58; \hspace{10ex} y_c=0.0461; \hspace{10ex} \Delta \chi^2 & = & 3.78;
\end{eqnarray*}

ALEPH:
\begin{eqnarray*}
9.6816(x-x_a)^2-2554.0(x-x_a)(y-y_a)+287560(y-y_a)^2&=&1;\\
x_a=0.74; \hspace{10ex} y_a=0.0361; \hspace{10ex} \Delta \chi^2 & = & 3.52;
\end{eqnarray*}

OPAL:
\begin{eqnarray*}
19.4601(x-x_o)^2-3771.3(x-x_o)(y-y_o)+379980(y-y_o)^2&=&1;\\
x_o=1.44; \hspace{10ex} y_o=0.0414; \hspace{10ex} \Delta \chi^2 & = & 0.10;
\end{eqnarray*}

DELPHI:
\begin{eqnarray*}
14.1777(x-x_d)^2-2856.4(x-x_d)(y-y_d)+246120(y-y_d)^2&=&1;\\
x_d=1.22; \hspace{10ex} y_d=0.0378; \hspace{10ex} \Delta \chi^2 & = & 1.47; 
\end{eqnarray*}

CLEO+ALEPH+OPAL+DELPHI (without common theoretical normalization error;
divide all coefficients by 2.850 for ellipse with this error):
\begin{eqnarray*}
85.2008(x-\bar{x})^2-14356(x-\bar{x})(y-\bar{y})+1152453(y-\bar{y})^2&=&1;\\
\bar{x}=1.27; \hspace{10ex} \bar{y}=0.0399; \hspace{10ex}
\end{eqnarray*}

CLEO+ALEPH+OPAL+DELPHI+DECAYS (without common theoretical normalization
error; divide all coefficients by 3.514 for ellipse with this error):
\begin{eqnarray*}
1059.7(x-\tilde{x})^2-86781(x-\tilde{x})(y-\tilde{y})+
2524400(y-\tilde{y})^2 & = & 1;\\
\tilde{x}=1.52; \hspace{10ex} \tilde{y}=0.0415; \hspace{10ex}
\end{eqnarray*}

% Journal and other miscellaneous abbreviations for references
% Phys. Rev. D format
\def \ajp#1#2#3{Am.\ J. Phys.\ {\bf#1}, #2 (#3)}
\def \apny#1#2#3{Ann.\ Phys.\ (N.Y.) {\bf#1}, #2 (#3)}
\def \app#1#2#3{Acta Phys.\ Polonica {\bf#1}, #2 (#3)}
\def \arnps#1#2#3{Ann.\ Rev.\ Nucl.\ Part.\ Sci.\ {\bf#1}, #2 (#3)}
\def \art{and references therein}
\def \cmts#1#2#3{Comments on Nucl.\ Part.\ Phys.\ {\bf#1}, #2 (#3)}
\def \cn{Collaboration}
\def \cp89{{\it CP Violation,} edited by C. Jarlskog (World Scientific,
Singapore, 1989)}
\def \efi{Enrico Fermi Institute Report No.\ }
\def \epjc#1#2#3{Eur.\ Phys.\ J. C {\bf#1}, #2 (#3)}
\def \f79{{\it Proceedings of the 1979 International Symposium on Lepton and
Photon Interactions at High Energies,} Fermilab, August 23-29, 1979, ed. by
T. B. W. Kirk and H. D. I. Abarbanel (Fermi National Accelerator Laboratory,
Batavia, IL, 1979}
\def \hb87{{\it Proceeding of the 1987 International Symposium on Lepton and
Photon Interactions at High Energies,} Hamburg, 1987, ed. by W. Bartel
and R. R\"uckl (Nucl.\ Phys.\ B, Proc.\ Suppl., vol.\ 3) (North-Holland,
Amsterdam, 1988)}
\def \ib{{\it ibid.}~}
\def \ibj#1#2#3{~{\bf#1}, #2 (#3)}
\def \ichep72{{\it Proceedings of the XVI International Conference on High
Energy Physics}, Chicago and Batavia, Illinois, Sept. 6 -- 13, 1972,
edited by J. D. Jackson, A. Roberts, and R. Donaldson (Fermilab, Batavia,
IL, 1972)}
\def \ijmpa#1#2#3{Int.\ J.\ Mod.\ Phys.\ A {\bf#1}, #2 (#3)}
\def \ite{{\it et al.}}
\def \jhep#1#2#3{JHEP {\bf#1}, #2 (#3)}
\def \jpb#1#2#3{J.\ Phys.\ B {\bf#1}, #2 (#3)}
\def \lg{{\it Proceedings of the XIXth International Symposium on
Lepton and Photon Interactions,} Stanford, California, August 9--14 1999,
edited by J. Jaros and M. Peskin (World Scientific, Singapore, 2000)}
\def \lkl87{{\it Selected Topics in Electroweak Interactions} (Proceedings of
the Second Lake Louise Institute on New Frontiers in Particle Physics, 15 --
21 February, 1987), edited by J. M. Cameron \ite~(World Scientific, Singapore,
1987)}
\def \kdvs#1#2#3{{Kong.\ Danske Vid.\ Selsk., Matt-fys.\ Medd.} {\bf #1},
No.\ #2 (#3)}
\def \ky85{{\it Proceedings of the International Symposium on Lepton and
Photon Interactions at High Energy,} Kyoto, Aug.~19-24, 1985, edited by M.
Konuma and K. Takahashi (Kyoto Univ., Kyoto, 1985)}
\def \mpla#1#2#3{Mod.\ Phys.\ Lett.\ A {\bf#1}, #2 (#3)}
\def \nat#1#2#3{Nature {\bf#1}, #2 (#3)}
\def \nc#1#2#3{Nuovo Cim.\ {\bf#1}, #2 (#3)}
\def \nima#1#2#3{Nucl.\ Instr.\ Meth. A {\bf#1}, #2 (#3)}
\def \np#1#2#3{Nucl.\ Phys.\ {\bf#1}, #2 (#3)}
\def \npbps#1#2#3{Nucl.\ Phys.\ B Proc.\ Suppl.\ {\bf#1}, #2 (#3)}
\def \os{XXX International Conference on High Energy Physics, Osaka, Japan,
July 27 -- August 2, 2000}
\def \PDG{Particle Data Group, D. E. Groom \ite, \epjc{15}{1}{2000}}
\def \pisma#1#2#3#4{Pis'ma Zh.\ Eksp.\ Teor.\ Fiz.\ {\bf#1}, #2 (#3) [JETP
Lett.\ {\bf#1}, #4 (#3)]}
\def \pl#1#2#3{Phys.\ Lett.\ {\bf#1}, #2 (#3)}
\def \pla#1#2#3{Phys.\ Lett.\ A {\bf#1}, #2 (#3)}
\def \plb#1#2#3{Phys.\ Lett.\ B {\bf#1}, #2 (#3)}
\def \pr#1#2#3{Phys.\ Rev.\ {\bf#1}, #2 (#3)}
\def \prc#1#2#3{Phys.\ Rev.\ C {\bf#1}, #2 (#3)}
\def \prd#1#2#3{Phys.\ Rev.\ D {\bf#1}, #2 (#3)}
\def \prl#1#2#3{Phys.\ Rev.\ Lett.\ {\bf#1}, #2 (#3)}
\def \prp#1#2#3{Phys.\ Rep.\ {\bf#1}, #2 (#3)}
\def \ptp#1#2#3{Prog.\ Theor.\ Phys.\ {\bf#1}, #2 (#3)}
\def \rmp#1#2#3{Rev.\ Mod.\ Phys.\ {\bf#1}, #2 (#3)}
\def \rp#1{~~~~~\ldots\ldots{\rm rp~}{#1}~~~~~}
\def \si90{25th International Conference on High Energy Physics, Singapore,
Aug. 2-8, 1990}
\def \slc87{{\it Proceedings of the Salt Lake City Meeting} (Division of
Particles and Fields, American Physical Society, Salt Lake City, Utah, 1987),
ed. by C. DeTar and J. S. Ball (World Scientific, Singapore, 1987)}
\def \slac89{{\it Proceedings of the XIVth International Symposium on
Lepton and Photon Interactions,} Stanford, California, 1989, edited by M.
Riordan (World Scientific, Singapore, 1990)}
\def \smass82{{\it Proceedings of the 1982 DPF Summer Study on Elementary
Particle Physics and Future Facilities}, Snowmass, Colorado, edited by R.
Donaldson, R. Gustafson, and F. Paige (World Scientific, Singapore, 1982)}
\def \smass90{{\it Research Directions for the Decade} (Proceedings of the
1990 Summer Study on High Energy Physics, June 25--July 13, Snowmass, Colorado),
edited by E. L. Berger (World Scientific, Singapore, 1992)}
\def \tasi{{\it Testing the Standard Model} (Proceedings of the 1990
Theoretical Advanced Study Institute in Elementary Particle Physics, Boulder,
Colorado, 3--27 June, 1990), edited by M. Cveti\v{c} and P. Langacker
(World Scientific, Singapore, 1991)}
\def \yaf#1#2#3#4{Yad.\ Fiz.\ {\bf#1}, #2 (#3) [Sov.\ J.\ Nucl.\ Phys.\
{\bf #1}, #4 (#3)]}
\def \zhetf#1#2#3#4#5#6{Zh.\ Eksp.\ Teor.\ Fiz.\ {\bf #1}, #2 (#3) [Sov.\
Phys.\ - JETP {\bf #4}, #5 (#6)]}
\def \zpc#1#2#3{Zeit.\ Phys.\ C {\bf#1}, #2 (#3)}
\def \zpd#1#2#3{Zeit.\ Phys.\ D {\bf#1}, #2 (#3)}

\end{document}